\documentclass[12pt, twoside, nofootinbib]{article}
\newcommand{\mystyles}{}
\newcommand{\mybib}{}

\usepackage{amsmath, amsthm, amssymb} % Nice equation type seting
\usepackage{hyperref}
\usepackage{epsfig}         % For including postscript figures;
\usepackage{graphicx}       % For including postscript figures; pdf, jpg graphics
\usepackage{\mystyles footnpag}  % For resetting footnote on every page.
\usepackage{latexsym}          % For having \Box defined.
\usepackage{\mystyles quotes}	 % automatically translates \" to `` or ''
\usepackage{\mybib natbib}% for bibliograpphy like in astronomy
\bibpunct{(}{)}{;}{a}{}{,} 	% to follow the A&A style

\usepackage{fancyhdr}
\begin{document}

\title{{\LARGE \textbf{The DTFE public software}}\\ {\large -The Delaunay Tessellation Field Estimator code -}\\ \vspace{.5cm} }

\author{\textbf{Marius Cautun}\thanks{cautun@astro.rug.nl}, \textbf{Rien van de Weygaert}\\
 Kapteyn Astronomical Institute,\\
University of Groningen, Netherlands \\
}

\maketitle
\thispagestyle{empty}

\pagenumbering{arabic}
\pagestyle{fancy}
\fancyhead{} % clear all header fields
\fancyhead[LE]{\bfseries M. Cautun, R. van de Weygaert}
\fancyhead[RO]{\bfseries The DTFE public software}
\fancyfoot{} % clear all footer fields
\fancyfoot[LE,RO]{\thepage}

\begin{abstract}
We present the DTFE public software, a code for reconstructing fields from a discrete set of samples/measurements using the maximum of information contained in the point distribution. The code is written in C++ using the \textbf{CGAL}\footnote{Computational Geometry Algorithms Library, http://www.cgal.org} library and is parallelized using OpenMP. The software was designed for the analysis of cosmological data but can be used in other fields where one must interpolate quantities given at a discrete point set. The software comes with a wide suite of options to facilitate the analysis of 2- and 3-dimensional data and of both numerical simulations and galaxy redshift surveys. For comparison purposes, the code also implements the TSC and SPH grid interpolation methods. The code comes with an extensive user guide detailing the program options, examples and the inner workings of the code. The DTFE public software and further information can be found at \url{http://www.astro.rug.nl/~voronoi/DTFE/dtfe.html}.

\end{abstract}

\section{Introduction: the DTFE method}
The Delaunay Tessellation Field Estimator (from now on DTFE) \citep[see][]{2000A&A...363L..29S,2009LNP...665..291V,Cautun2011} represents the natural method of reconstructing from a discrete set of samples/measurements a volume-covering and continuous density and intensity fields using the maximum of information contained in the point distribution. This interpolation method works for any scalar or vector fields (e.g. velocity, temperature, pressure) that are defined at the positions of a discrete point set. Moreover, if the point distribution traces the underlying density field, DTFE offers a local method in determining the density at each point and accordingly also in the whole volume.
\begin{figure}[h]
	\begin{center}
	\includegraphics[width=.7\linewidth]{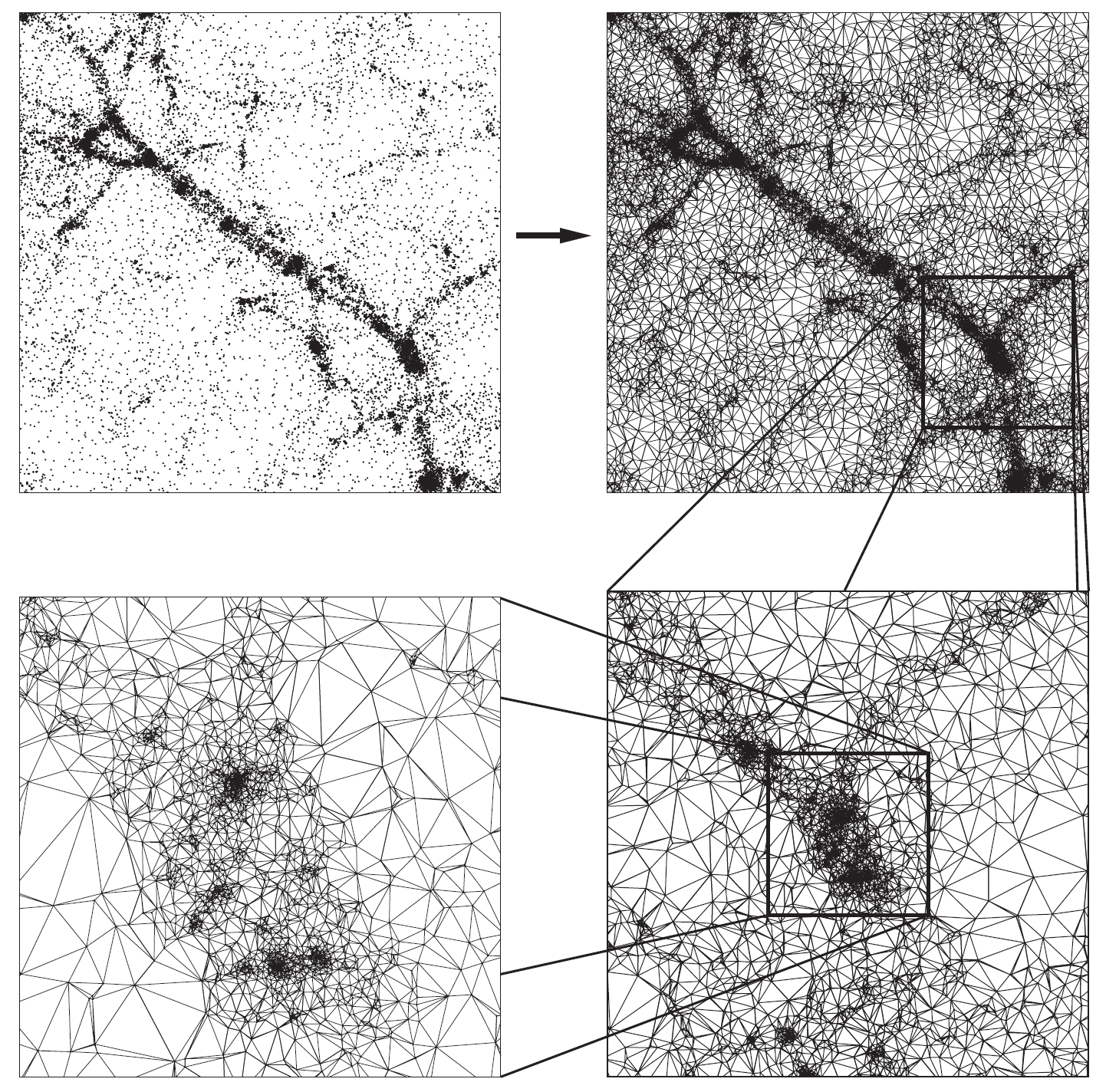}
	\caption{An illustration of the 2D Delaunay tessellation of a set of particles from a cosmological numerical simulation. Courtesy: \citet{Schaap2007}.}
	\label{fig=DTFE_example}
	\end{center}
\end{figure}

The DTFE method was first developed by Willem Schaap and Rien van de Weygaert \citep{2000A&A...363L..29S} to be used on various astrophysical applications, but can also be used in other fields where one must interpolate using quantities given at a discrete point set. The DTFE method is especially suitable for astrophysical data due to the following reasons:
\begin{enumerate}
	\item Preserves the multi-scale character of the point distribution. This is the case in numerical simulations of large scale structure (from now on LSS) where the density varies over more than 6 orders of magnitude and for a lesser extent for galaxy redshift surveys.
	\item Preserves the local geometry of the point distribution. This is important in recovering sharp and anisotropic features like the different components of the LSS (i.e. clusters, filaments, walls and voids).
	\item DTFE does not depend on user defined parameters or choices.
	\item The interpolated fields are volume weighted (versus mass weighted quantities in most other interpolation schemes). This can have a significant effect especially when comparing with analytical predictions which are volume weighted \citep[see][]{1996MNRAS.279..693B}.
\end{enumerate}
The first two points can be easily seen in Fig. \ref{fig=DTFE_example}, where starting from the point distribution (upper left insert), one constructs the Delaunay tessellation (upper right insert) and than zooms on a high density region (lower right and left inserts).

\begin{figure}[h]
	\begin{center}
	\includegraphics[width=.99\linewidth]{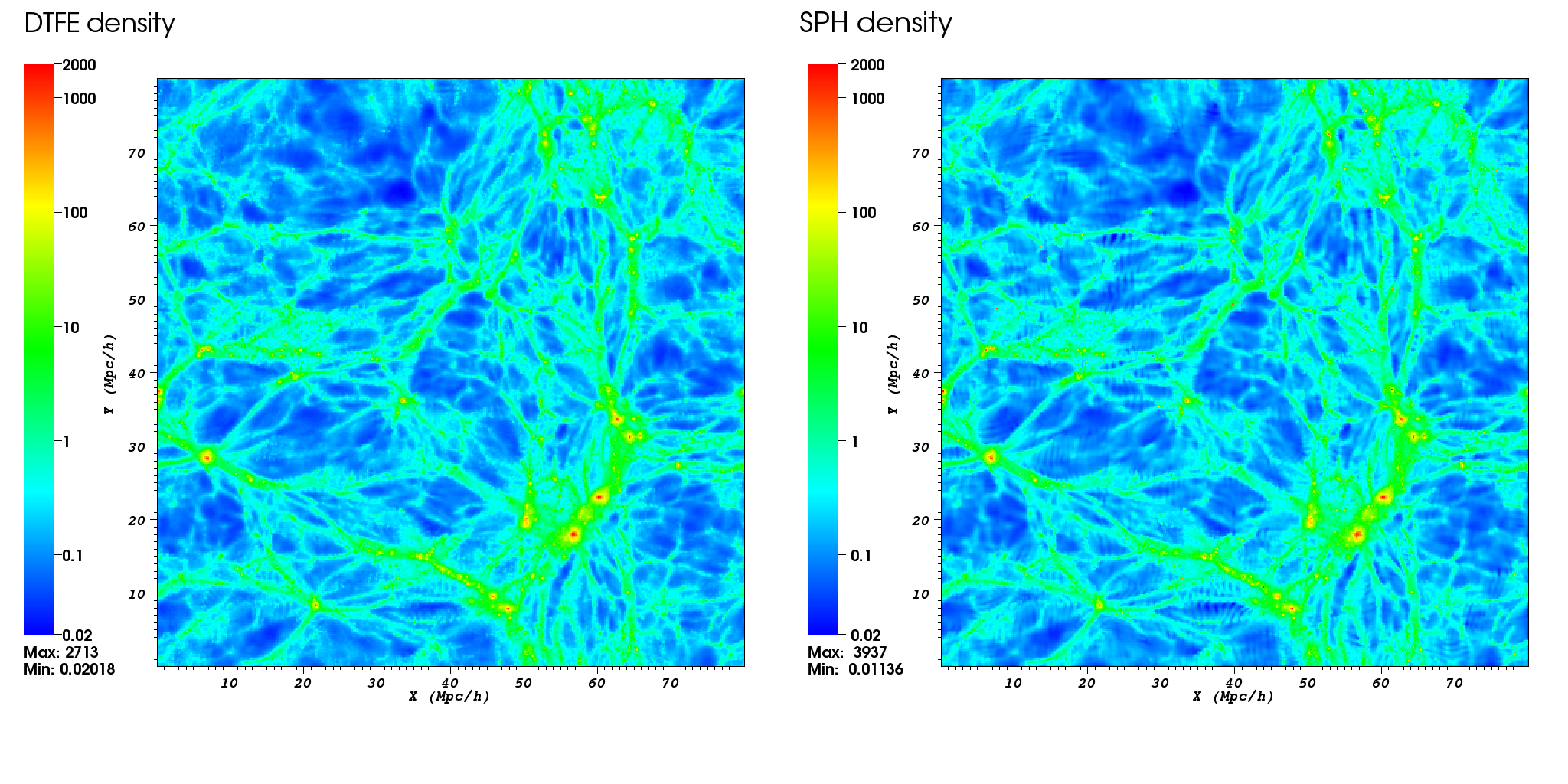}
	\caption{A comparison between the DTFE (left panel) and SPH (right panel) densities through a $80\times80\times2$ Mpc/h slice in an N-body simulation. Both density fields are represented on a logarithmic scale. The SPH density was computed using an adaptive smoothing length that involves the closest 40 neighbors.}
	\label{fig=DTFE_density}
	\end{center}
\end{figure}
\begin{figure}[h]
	\begin{center}
	\includegraphics[width=.99\linewidth]{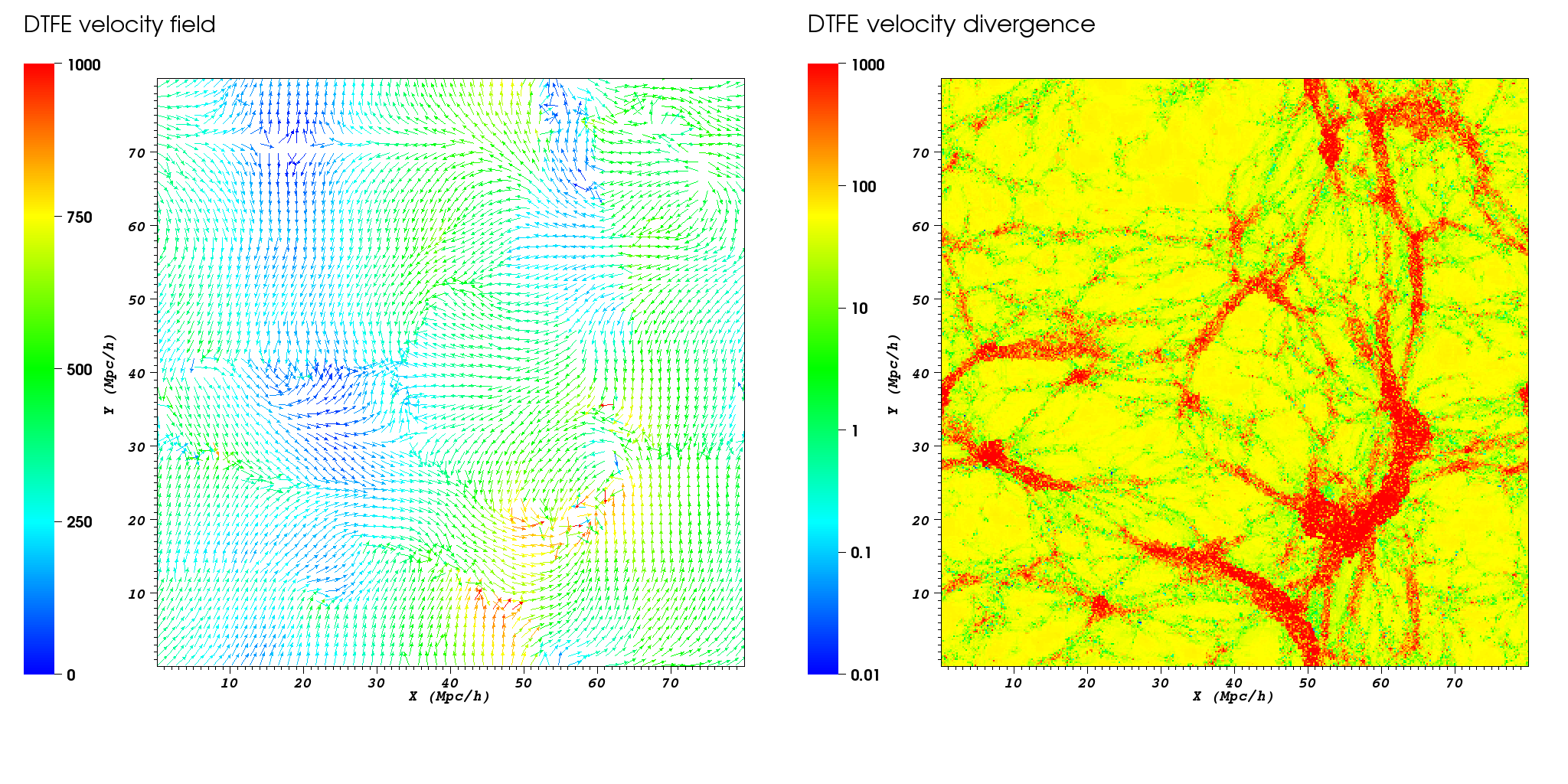}
	\caption{A map of the DTFE computed velocity flow (left panel) and velocity divergence (right panel) corresponding to the density field shown in Fig. \ref{fig=DTFE_density}. The left panel shows the velocity vectors colored according to the velocity magnitude. The right panel show the absolute value of the velocity divergence such that it can be shown on a logarithmic scale.}
	\label{fig=DTFE_velocity}
	\end{center}
\end{figure}

\section{The DTFE public software}
The code accompanying this document is a C++ implementation of the DTFE method of interpolating from a discrete point set (in 2 or 3 dimensions) to a grid. The software was designed using a modular philosophy and with a wide set of features that can easily be selected using the different program options. These features were constructed to analyze the data using multiple methods and to manipulate and split the data such that it can be useful in dealing with a wide variety of problems. The DTFE code is also written using OpenMP directives which allow it to run in parallel on shared-memory architectures. The software can be used as a standalone program or as an external library.\\

The code comes with complete documentation and with a multitude of examples that detail the program features. Moreover, a help desk is available for information and assistance for troubleshooting problems. The DTFE software can be downloaded from \url{http://www.astro.rug.nl/~voronoi/DTFE/dtfe.html}. The same address contains the user guide as well as help and contact forms.

\section{DTFE software features}
The DTFE software has a wide range of features designed to facilitate the work with cosmological data, both from numerical simulations and from observations:
\begin{enumerate}
	\item[$\bullet$] Works in both 2 and 3 spatial dimensions.
  \item[$\bullet$] Interpolates the fields to three different types of grids:
  	\begin{enumerate}
			\item[1.] Regular rectangular and cuboid grid --- useful for numerical simulation.
			\item[2.] Redshift cone (spherical coordinates) grid --- useful for galaxy redshift survey or for simulating observations.
			\item[3.] User given sampling points --- can describe any complex or non-regular sampling geometry
		\end{enumerate}
  \item[$\bullet$] Uses the point distribution itself to compute the density and interpolates the result to grid.
  \item[$\bullet$] Each sample point has a weight associated to it to represent multiple resolution N-body simulations and observational biases for galaxy redshift surveys.
  \item[$\bullet$] Interpolates the velocity, velocity gradient, velocity divergence, velocity shear and velocity vorticity (see Fig. \ref{fig=DTFE_velocity} for an example of the DTFE velocity and velocity divergence fields).
  \item[$\bullet$] Interpolates any additional number of fields and their gradients to grid.
  \item[$\bullet$] Periodic boundary conditions.
  \item[$\bullet$] Zoom in option for regions of interest.
  \item[$\bullet$] Splitting the full data in smaller computational chunks when dealing with limited CPU resources.
  \item[$\bullet$] The computation can be distributed in parallel on shared-memory architectures.
  \item[$\bullet$] For comparison purposes, the software comes also with the TSC\citep{1981csup.book.....H} and the SPH\citep{1992ARA&A..30..543M} interpolation methods (see Fig. \ref{fig=DTFE_density} for a comparison of DTFE and SPH density fields).
  \item[$\bullet$] Can return the Delaunay tessellation of the given point set.
  \item[$\bullet$] Easy change of input/output data format.
  \item[$\bullet$] Easy to use as an external library.
  \item[$\bullet$] Extensive documentation of each feature.
\end{enumerate}

\section{DTFE software license}
DTFE is free software, distributed under the \href{http://www.gnu.org/copyleft/gpl.html}{GNU General Public License}. This implies that you may freely distribute and copy the software. You may also modify it as you wish, and distribute these modified versions as long as you indicate prominently any changes you made in the original code, and as long as you leave the copyright notices, and the no-warranty notice intact. Please read the General Public License for more details. Note that the authors retain their copyright on the code.\\

If you use the DTFE software for scientific work, we kindly ask you to reference the DTFE method and code papers: \citet{2000A&A...363L..29S}, \citet{2009LNP...665..291V} and \citet{Cautun2011}.\\

\noindent{\small \textbf{Acknowledgments.} The authors acknowledges the work of Erwin Platen who wrote a basic version of the DTFE density interpolation code which was used as the starting point of this work. We are also grateful for discussions with Patrick Bos, Pratyush Pranav, Miguel Aragon-Calvo and Johan Hidding whose suggestions shaped the form and features of the software. We are also grateful to Monique Teilland, Manuel Caroli and Gert Vegter for their work in developing the CGAL library as well as their help and support with the CGAL library. We thank Willem Schaap for his work in developing the DTFE method as a mature analysis tool for studying cosmological data. A special acknowledgment goes to Bernard Jones, whose continuing involvement and encouragement during the various stages of DTFE development have been instrumental.}

\newcommand{\araa}{Annual Review of Astronomy and Astrophysics}
\newcommand{\apj}{The Astrophysical Journal}
\newcommand{\apjs}{The Astrophysical Journal Supplement}
\newcommand{\mnras}{M.N.R.A.S.}
\newcommand{\aap}{Astronomy and Astrophysics}
\nocite{cgal}

\bibliographystyle{\mybib aa}
\bibliography{bibliography}

\begin{thebibliography}{8}
\expandafter\ifx\csname natexlab\endcsname\relax\def\natexlab#1{#1}\fi

\bibitem[{cga(????)}]{cgal}
\textsc{Cgal}, {C}omputational {G}eometry {A}lgorithms {L}ibrary, 2D and
  3D Triangulation Packages, \url{http://www.cgal.org}

\bibitem[{{Bernardeau} \& {van de Weygaert}(1996)}]{1996MNRAS.279..693B}
{Bernardeau}, F. \& {van de Weygaert}, R. 1996, \mnras, 279, 693

\bibitem[{{Cautun} {et~al.}(2011){Cautun}, {Schaap}, \& {van de
  Weygaert}}]{Cautun2011}
{Cautun}, M., {Schaap}, W., \& {van de Weygaert}, R. 2011, \mnras, in prep.

\bibitem[{{Hockney} \& {Eastwood}(1981)}]{1981csup.book.....H}
{Hockney}, R.~W. \& {Eastwood}, J.~W. 1981, {Computer Simulation Using
  Particles}, ed. {Hockney, R.~W.~\& Eastwood, J.~W.}

\bibitem[{{Monaghan}(1992)}]{1992ARA&A..30..543M}
{Monaghan}, J.~J. 1992, \araa, 30, 543

\bibitem[{{Schaap}(2007)}]{Schaap2007}
{Schaap}, W.~E. 2007, The Delaunay Tessellation Field Estimator, PhD thesis,
  University of Groningen

\bibitem[{{Schaap} \& {van de Weygaert}(2000)}]{2000A&A...363L..29S}
{Schaap}, W.~E. \& {van de Weygaert}, R. 2000, \aap, 363, L29

\bibitem[{{van de Weygaert} \& {Schaap}(2009)}]{2009LNP...665..291V}
{van de Weygaert}, R. \& {Schaap}, W. 2009, Lecture Notes in Physics, Berlin
  Springer Verlag, Vol. 665, {The Cosmic Web: Geometric Analysis}, ed.
  {V.~J.~Mart{\'{\i}}nez, E.~Saar, E.~Mart{\'{\i}}nez-Gonz{\'a}lez, \&
  M.-J.~Pons-Border{\'{\i}}a}, 291--413

\end{thebibliography}

\end{document}